\documentclass[conference]{IEEEtran}
\IEEEoverridecommandlockouts
\usepackage{cite}
\usepackage{amsmath,amssymb,amsfonts}
\usepackage{graphicx}
\usepackage{textcomp}
\usepackage{xcolor}
\usepackage{bm}
\usepackage{algorithm}
\usepackage{algpseudocode}
\usepackage{etoolbox}
\usepackage{colortbl}
\usepackage{booktabs}
\usepackage{multirow}
\def\BibTeX{{\rm B\kern-.05em{\sc i\kern-.025em b}\kern-.08em
    T\kern-.1667em\lower.7ex\hbox{E}\kern-.125emX}}
\begin{document}

\title{Vulnerability of Building Energy Management against Targeted False Data Injection Attacks: Model Predictive Control vs. Proportional Integral}
\author{\IEEEauthorblockN{\textsuperscript{}  Xiaoyu Ge$^1$, Kamelia Norouzi$^1$, Faegheh Moazeni$^1$, Mirel Sehic$^2$, Javad Khazaei$^{1,*}$, Parv Venkitasubramaniam$^1$, \\and Rick Blum$^1$ }
\IEEEauthorblockA{\textit{$^1$ Rossin college of engineering and applied science, $^2$ Honeywell Inc.} \\
\textit{Lehigh University},
Bethlehem PA, USA \\
Emails: \textit{(xig620, kan423,moazeni, khazaei, pav309, rb0f)@lehigh.edu, mirel.sehic@honeywell.com}}

\thanks{This research was financed (in part) by a grant from the Commonwealth of Pennsylvania,
Department of Community and Economic Development, through the Pennsylvania Infrastructure
Technology Alliance (PITA). }
}
\IEEEpubidadjcol
\IEEEpubid{\begin{minipage}{\textwidth}\ \\[25pt] \centering
\color{blue}This work has been submitted to the IEEE for possible publication. Copyright may be transferred without notice, after which this version may no longer be accessible.
\end{minipage}}

\maketitle

\begin{abstract}
Cybersecurity in building energy management is crucial for protecting infrastructure, ensuring data integrity, and preventing unauthorized access or manipulation. This paper investigates the energy efficiency and cybersecurity of building energy management systems (BMS) against false data injection (FDI) attacks using proportional-integral (PI) and model predictive control (MPC) methods. Focusing on a commercial building model with five rooms, vulnerability of PI-based BMS and nonlinear MPC-based BMS against FDIs on sensors and actuators is studied. The study aims to assess the effectiveness of these control strategies in maintaining system performance and lifespan, highlighting the potential of MPC in enhancing system resilience against cyber threats. Our case studies demonstrate that even a short term FDIA can cause a 12\% reduction in lifetime of a heat-pump under an MPC controller, and cause a near thirty-fold overuse of flow valves under a PI controller. 
\end{abstract}


\section{Introduction}
Smart buildings have become an integral part of our life, largely thanks to their automated energy management system (BMS). Central to this system are heating, ventilation, and air conditioning (HVAC) systems that are crucial for ensuring a comfortable and secure environment for residents.  Notably, in the European building sector, HVAC systems account for approximately 40 to 60\% of energy usage, while this figure surpasses 50\% in the United States \cite{solano2021hvac}. As such, the HVAC sector has garnered significant attention for its potential for energy savings and sustainable building practices. The primary objective is to maintain a pleasant indoor environment and effective ventilation while minimizing energy usage, necessitating precise control of temperature and airflow rates  \cite{lamnabhi2017systems}. \par In the quest for enhanced building energy efficiency, various control strategies have been explored. Beyond traditional linear methods such as proportional integral (PI) controllers, which face challenges due to parameter variability and external disturbances \cite{xie2022ga,hussain2018internal,gupta2016controller,jiang2022building}, newer approaches such as Fuzzy logic \cite{fzy} and neural networks \cite{nn} have also been investigated. However, the advent of model predictive control (MPC) has marked a significant advancement in this field. By leveraging a predictive model of the system, MPC has demonstrated substantial energy savings in real-life applications, making it an increasingly popular choice for building BMS \cite{chen2022mbrl,ostadijafari2020tube,valenzuela2019closed}. \par  With the increased cyber-connectivity of BMS assets, a growing concern is vulnerabilities such as cyber-attacks, where attackers manipulate the measurements to impact the performance and security of BMS and their controllers \cite{sheikh2019cyber,ramos2023lstm}. While significant research has been dedicated to security of smart grids against cyberattacks \cite{reda2022comprehensive}, cybersecurity of building energy management systems has not been thoroughly investigated. For example, in \cite{cyb1}, cybersecurity of a smart home energy management under random false data injection (FDI) attacks its impact on increased charging and discharging cycles of home battery energy storage was investigated. In \cite{cyb2}, a game theoretic approach for home energy management under FDI attacks was proposed to identify random FDI attacks and schedule the home assets using game-theory-based scheduling. Existing research on cybersecurity of buildings has mainly focused on random FDI attacks  and to the best of our knowledge, no study has explored the vulnerability of BMS against targeted FDI attacks on lifetime of building assets. In addition, given a mix of PI-based and MPC-based BMS currently in operation, vulnerability of these control techniques against FDI attacks is unknown. 
\par To address these knowledge gaps, this paper aims at assessing the vulnerability of building energy management system against targeted FDI attacks. The main contribution of our study is the comparative assessment and analysis of targeted FDI attacks on both PI- and MPC-based controllers within an HVAC framework, assessing their performance and resilience under varied attack scenarios. The proposed approach lays a solid foundation to examine the resilience of BMS controllers and their robustness against potential cyber threats and their overall effectiveness in maintaining secure building operation.

The paper is organized as follows: Section II introduces a commercial building model. Section III discusses the cyberattack model and building control frameworks. Section IV presents case studies  and the paper concludes in Section V.
 
\section{Methodology}
\begin{figure}[t!]
\centering 
\includegraphics[width=3.5in]{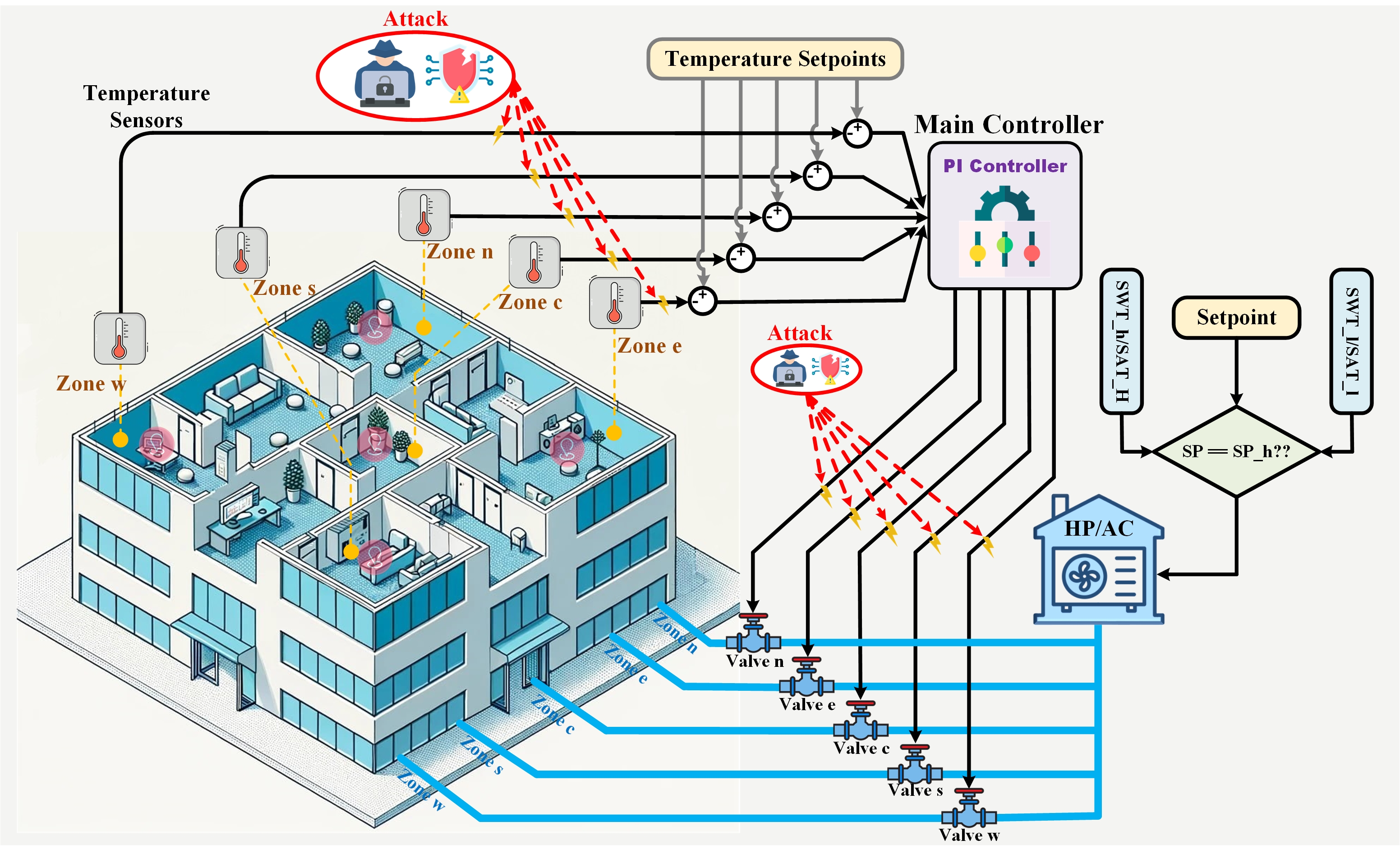}\vspace{-0.15in}
\caption{Building and PI control diagram }
\label{fig:PI}
\end{figure}

\begin{figure}[t!]
\centering 
\includegraphics[width=3.5in]{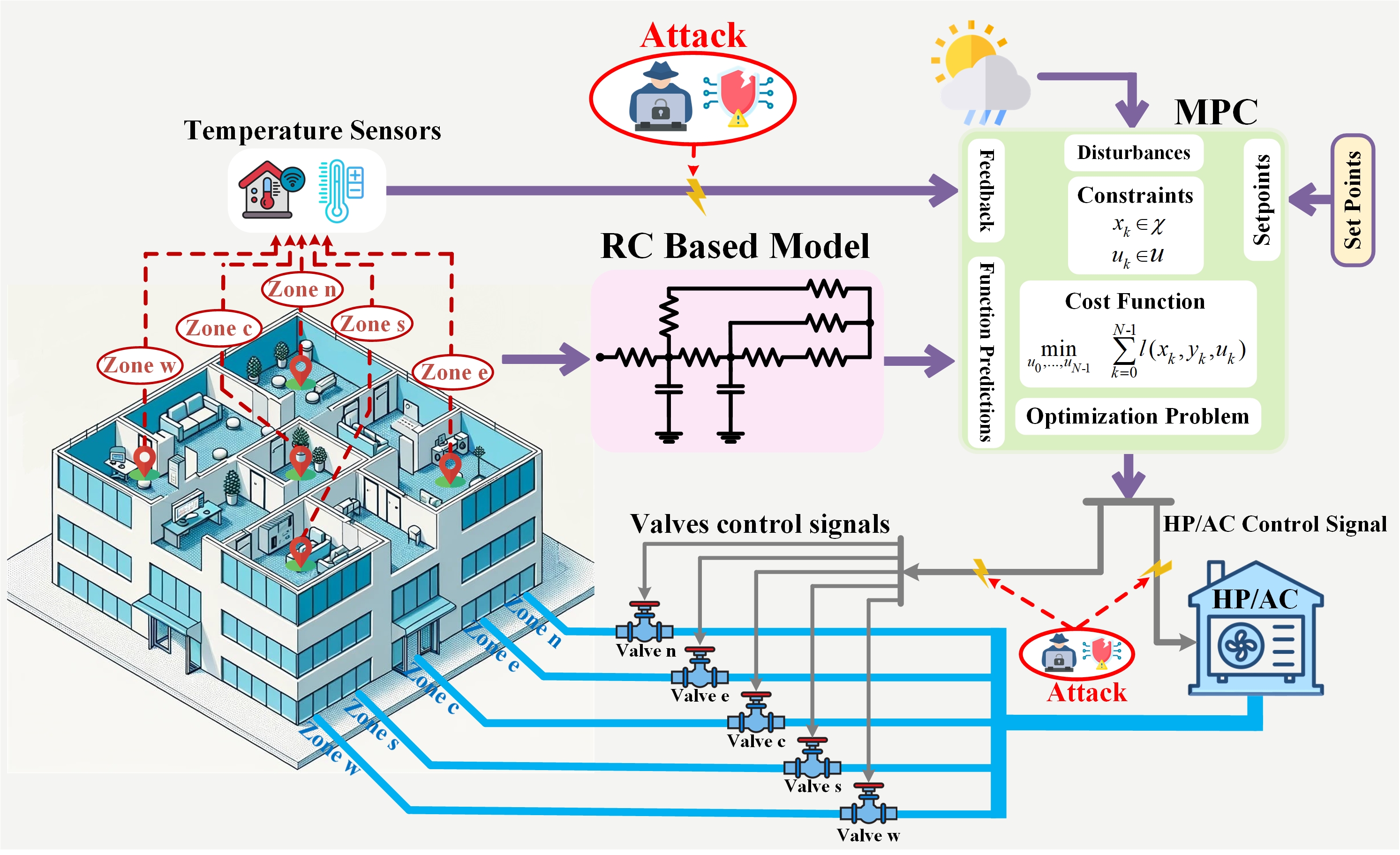}\vspace{-0.15in}
\caption{Building and MPC control diagram }
\label{fig:mpc}
\end{figure}

\begin{figure}[t!]
    \centering
    \includegraphics[scale=0.45]{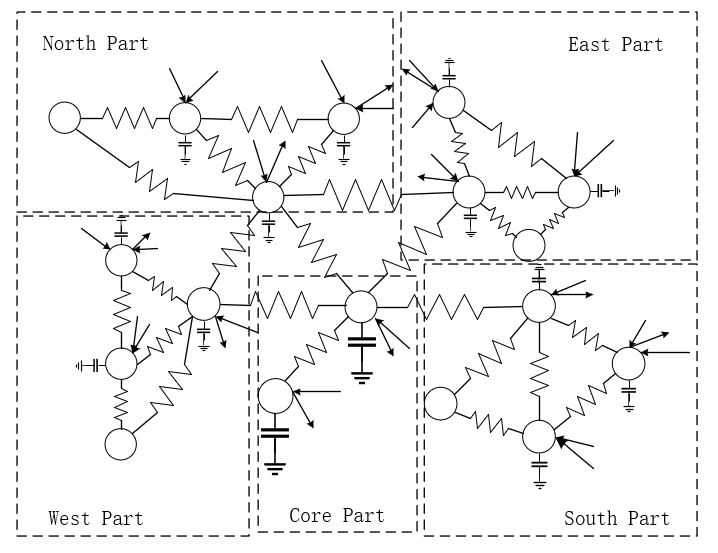}\vspace{-0.15in}
    \caption{RC Building Model}
    \label{RC_Model}
\end{figure}

A commercial building with a heat pump and two control systems including MPC and PI are considered. The building model can capture the temperature dynamics for a heat pump control.  The two main systems are shown in Fig.~\ref{fig:PI} and Fig.~\ref{fig:mpc} for PI-based BMS, and MPC-based BMS. In this paper, we consider a commercial building model with three thermal zones and five rooms. To represent the building model in terms of differential equations suitable for control design purposes, an RC model was adopted from \cite{zanetti2023performance} (see Fig.~\ref{RC_Model}).  

The thermal zones were developed as a network of three nodes, representing the dynamics of the air temperature of the zone, the floor temperature dynamics, and the dynamics of the external walls, which are expressed using the following differential equations:
\begin{align}
C_{r,x}&\dot{T}_{r,x}=R_{rf,x}(T_{f,x}-T_{r,x})+R_{rr}(T_{r,c}-T_{r,x})\notag \\&+R_{r-r}(T_{r,k}-T_{r,x})+R_{r-r}(T_{r,k}-T_{r,x})\notag \\&+R_{wr,x}(T_{x,w-T_{r,x}})+R_{er}(T_{e}-T_{r,x})\notag \\&+(1-b)\phi_{ig}+Q_{c,x}
\label{eql_room}
\end{align}
\begin{align}
C_{f,x}&\dot{T}_{f,x}=R_{rf,x}(T_{r,x}-T_{f,x})+c\phi_sA_{win,x}\notag \\&+R_{wf,x}(T_{w,x}-T_{f,x})+Q_{h,x}
\label{eql_floor}
\end{align}
\begin{align}
C_{w,x}&\dot{T}_{w,x}=R_{rw,x}(T_{r,x}-T_{w,x})+b\phi_{ig}\notag \\&+R_{wf,x}(T_{f,x}-T_{w,x})+R_{er}(T_e-T_{w,x})\notag \\&+a\phi_sA_{wall,x}
\label{eql_wall}
\end{align}
The parameters of the model with their definitions can be found in Table. \ref{tab:var}. 
\begin{table}[t!]
\caption{Variables involved in the building model}
\label{tab:var}
\begin{tabular}{ll}
\toprule[2pt]
\hline
$T_{r}$, $T_{f}$, $T_{w}$ & Room, floor, and wall temperature \\
$T_e$, $T_{sw}$, $T_{sa}$ & Weather, supply water, and supply air temperature \\ 
$C_{r}$ & Room air heat capacity \\
$C_{f}$ & Floor heat capacity \\
$C_{w}$ & External wall heat capacity \\
$C_{wt}$ & Water heat capacity \\
$\hat{m}$ & nominal value of water flow \\
$R_{rf}$ & Thermal resistance between the room and floor \\
$R_{wf}$ & Thermal resistance between the external wall and floor \\
$R_{rr}$ & Thermal resistance between two adjacent room \\ 
$R_{er}$ & Thermal resistance between the external weather and room \\
$\phi_{ig}$ & Internal gains \\
$\phi_{s}$ & Solar radiation \\
$A_{wall}$ & Wall area \\
$A_{win}$ & Window area \\
$V$ & Valve state \\\hline
\bottomrule[2pt]
\end{tabular}
\end{table}
Since the center room is located in the middle section of the building, disconnected from the external walls,  the dynamics of the center room are excluded.  In addition, $x$ is the index of the targeted room, $k$ is the index of the adjacent room, $Q_{c,x}$ represents the cooling flow due to the cooling system, $Q_{h,x}={H}_{in,x}-{H}_{out,x}$ is the heating flow due to the floor heating system, and ${H}_{in,x}-{H}_{out,x}=\hat{m}c_{wt}(T_{in}-T_{out})=\hat{m}c_w(1-w_f)(T_{sw}-T_f)$.

\section{Control framework}
\subsection{Baseline PI controller}
For the control baseline, we consider the rule-based PI controller. The controller operates in two different models of cooling and heating. The supply water temperature, supply air temperature, and valve states in each room are regulated through the controller to make the room temperature track the setpoint temperature. When the divergence between the room temperature and setpoint exceeds the tolerance of the comfortable range, the PI controller will adjust the supply water temperature or supply air temperature, which have two different setting levels. If the divergence remains significant, adjustments to valve settings in individual rooms will be made to regulate the flow rate. This modulation, combined with supply water (air) temperature, aims to minimize deviations between room temperature and setpoints. The control architecture is depicted in Fig. \ref{fig:PI}.
\subsection{Nonlinear Model Predictive Control}
Model predictive control is an advanced control method typically consisting of four key components: a predictive model, a cost function, constraints on variables, and an optimization algorithm that uses the predictive model outputs to minimize the cost under the constraints. The building thermal dynamics Eq.s (\ref{eql_room})-(\ref{eql_wall}) are presented as:
\begin{equation}
    \dot{\bm{x}} = \mathbf{f}(\bm{x}(t),\bm{u}(t),\bm{d}(t)) \label{ssmvdc}\vspace{-0.2cm}
\end{equation}
where  $\bm{x} \in \mathbb{R}^{n_x} = [T_{r,c} \ T_{f,c} \ T_{r,w} \ T_{f,w} \ T_{w,w} \ T_{r,e} \ T_{f,e} \ T_{w,e} \ T_{r,s}$ $\ T_{f,s} \ T_{w,s} \ T_{r,n} \ T_{f,n} \ T_{w,n}]^T$ is the state variable vector. $\bm{u} \in \mathbb{R}^{n_u} = [T_{SW} \ T_{SA} \ V_{c}\ V_{w}\ V_{e}\ V_{s}\ V_{n}]^T$ is the control variable vector and $\bm{d} \in \mathbb{R}^{n_d} = [T_{e} \ \phi_{s} \ \phi_{ig} ]^T$ represents the measured disturbance vector. Furthermore, this dynamic model over the prediction horizon can be written as: 
\begin{subequations}\label{nmpc_pred}
\begin{align}
    \bm{x}(k+n+1) &= \mathbf{f}(\bm{x}(k+n),\bm{u}(k+n),\bm{d}(k+n))  \\ 
    \bm{y}(k+n+1)&= \mathbf{g}(\bm{x}(k+n))\\& \quad \quad \forall n \in \{0,\dots,N_p-1\} \\& \quad \quad \forall k \in \{k_0,\dots,T-1\} \label{simulation_time}
\end{align}
\end{subequations}
where $k$ represents the simulation time step, starting from $k_0$ and ending at $T-1$. $N_p$ is the prediction horizon, and $n$ is the sampling time step. The MPC is formulated by:
\begin{subequations} \label{eq:ocp}
\begin{align}  \min_{\mathbf{u},\mathbf{x}} & \ \ J(k)=\sum_{m=1}^{N_o-1}\sum_{j=0}^{N_p} \| w_{m,j}^{o}\left[\bm{y}^{ref}_m(k+n)-y_m(k+n)\right]\|^2 \nonumber \\ \label{c-objfun}\\
    \text{s.t.} \quad & \bm{x}(k+j+1) = \mathbf{f}(\bm{x}(k+j), \bm{u}(k+j),\bm{d}(k+j)) \label{c-dynamic} \\
    \quad & \bm{y}(k+j+1) = \mathbf{g}(\bm{x}(k+j)) \quad    j = 0, \dots, N_p-1  \\ 
    & \bm{u}(k+j) \in \mathcal{U}, \quad j = 0, 1, \dots, N_p-1 \label{c-ubound}\\ 
    & \bm{x}(k+j) \in \mathcal{X}, \quad j = 1, 2, \dots, N_p-1 \label{c-statebound}\\
    &  \bm{x}(k+j+1) - \bm{x}(k+j) \leq \Delta \bm{x}  \\
    &  \bm{u}(k+j+1) - \bm{u}(k+j) \leq \Delta \bm{u} \label{eq:bound} 
\end{align}
\end{subequations}
where $N_p \in \mathbb{N}^+$ is the prediction horizon, $k=kt_s$ is the current control interval, $w_{m,j}^{o}$ is the tuning weight for the $j$th output.
After each sampling time, the MPC controller receives the updated measurements from sensors, and then computes the optimal supply water/air temperature and valve states by minimizing the cost function. In the cost function, $N_o$ is the set of outputs, while $y_m^{ref}$ and $y_m$ are the setpoint and measurement for the $m$th output, respectively. In addition, several constraints ($\bm{y}(k+n+1)= \mathbf{g}(\bm{x}(k+n)$) are added to states and control variables (e.g., room temperature, floor temperature, supply water/air temperature). The algorithm for operation of MPC is shown in Algorithm \ref{alg:nmpc}.

\begin{algorithm}[t!]
\caption{Nonlinear MPC Controller (NMPC)}
\label{alg:nmpc}
\begin{algorithmic}[1]
\State \textbf{Input Parameters:} $N_p,n_x,n_u,n_d,c_{wt},\hat{m},T_{end}$
\State \textbf{Input Data:} $\mathbf{y}^{ref}, \mathbf{f}(\bm{x}(t),\bm{u}(t),\bm{d}(t))$
\State \textbf{Initialize} $\bm{x}_0 \in \mathbb{R}^{n_x},\bm{u}_0 \in \mathbb{R}^{n_u}, \bm{d}_0 \in \mathbb{R}^{n_d},k=0$
\For {$\mathrm{k=0 \to T_{end}-1}$} \Comment{Simulation time}
\For {$j=0$ to $N_p-1$}
\State \textbf{Define \& Discretize} state-space model, \eqref{ssmvdc}
\State \textbf{Construct} prediction model, \eqref{nmpc_pred}
\State \textbf{Formulate} $J(\bm{x}(k+j),\bm{u}(k+j))$, \eqref{c-objfun}
\EndFor
\State \textbf{Solve} $J(k+j)$ s.t. \eqref{c-dynamic}-\eqref{eq:bound}
\State \textbf{Extract} $[{\bm{u}^{\ast}(0|k),\hdots,\bm{u}^{\ast}{(N_p-1|k)}}]$.
\State \textbf{Apply} only $\bm{u}^{\ast}(0|k)$
\State \textbf{Measure} $\bm{x}(k+1|k)$ from \eqref{eql_room} to \eqref{eql_wall}
\State \textbf{Update} for $k+1$, $\bm{x}_0 = \bm{x}(k+1|k)$ and $\bm{u}_0 = \bm{u}^{\ast}(0|k)$ 
\EndFor
\end{algorithmic}
\end{algorithm}
\section{Cyberattack Model}
\begin{figure}[t!]
\centering 
\includegraphics[width=3in]{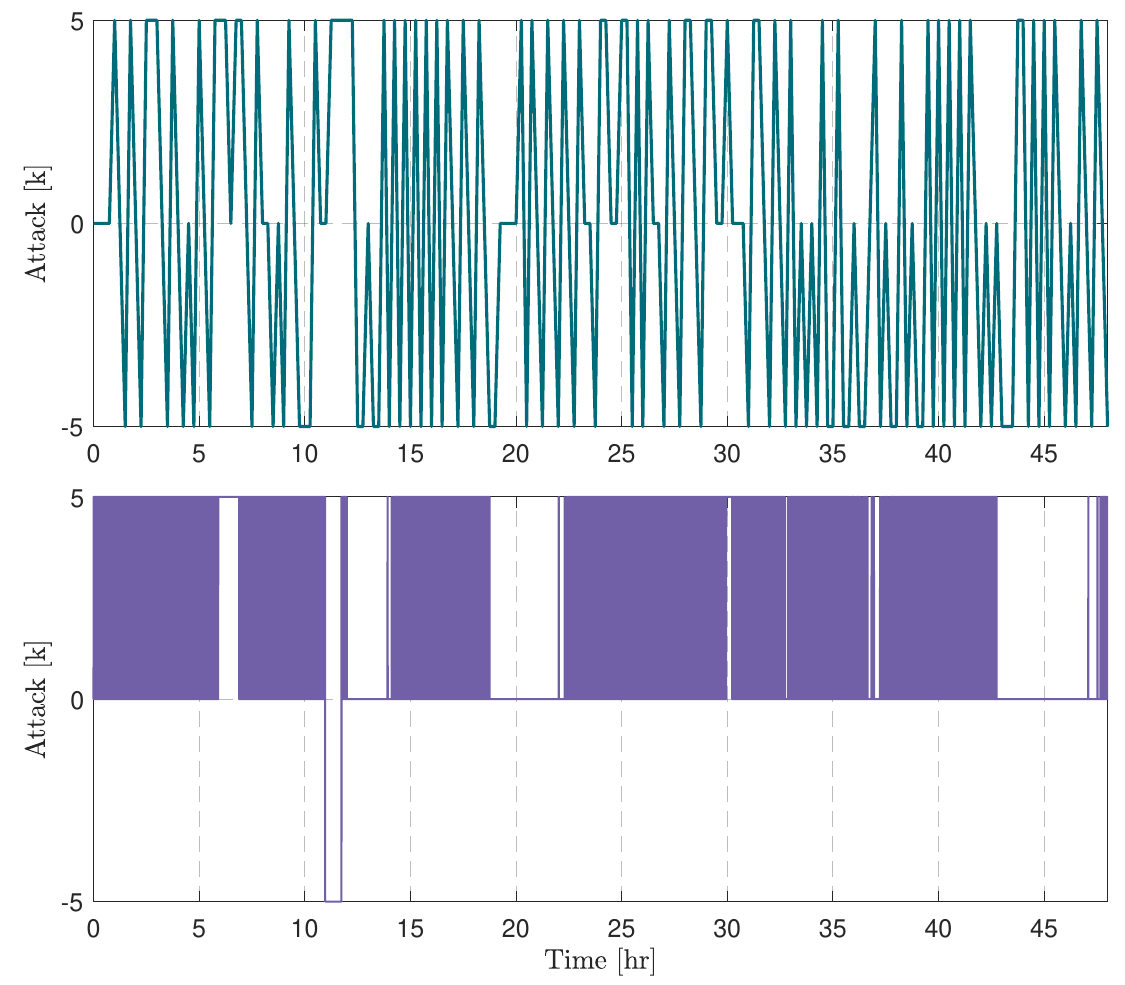}\vspace{-0.15in}
\caption{Obtained FDI attack for south room temperature sensor with MPC and PI controllers. }
\label{fig.attack}
\end{figure}

False data injection attack (FDIA) is one of the most common cyber-attacks, particularly in cyber-physical systems. FDI can cause the sensors to change the measurements before they are sent to the controller, and also can be applied on the actuators to change the control variables. 
The goal of this paper is to determine a targeted cyber-attack that would affect the HVAC system's lifetime. Building on this analysis, the attack strategy can be conceptualized as an optimization problem, aiming to maximize the reduction of the heat pump lifetime within defined constraints. The attack model is expressed as:
\begin{equation}
\begin{array}{rrclcl}
\displaystyle \min_{T_A} & \multicolumn{3}{l}{-\dfrac{\hat{m}c_{wt}(1-wf)V(T_{sw}-T_{f})(T_{r}+T_A-T_{e})}{(T_{r}+T_A)}} \label{obj}\\
\textrm{s.t.} & \mathrm{Eqs}. \ \ \eqref{eql_room}-\eqref{eql_wall}\\
&\underline{T}_A \leq T_A \leq \overline{T}_A   \\
\end{array}
\end{equation}
where the objective function in \eqref{obj} aims at maximizing the cost of running the heat pump system (i.e., minimizing the negative cost), which is also identical to maximizing the utilization of heat pump system during the day thus reducing its lifetime. For the lifetime evaluation, two criteria are selected: operation times of valves and heat pump consumption lifetime. The total power consumption of the heat pump which is used to evaluate the lifespan of the heat pump is formulated as:
\begin{equation}
    Q= \sum_{i=1}^{5} \frac{\hat{m}c_{wt}(1-wf)V_i(T_{sw}-T_{f,i})(T_{r,i}-T_{e})}{T_{r,i}}  \label{Eq:hppc}
\end{equation} which depends on floor temperature reading $T_f$, room temperature $T_r$, and external weather temperature reading $T_e$. The objective function \eqref{obj} is the modified version of \eqref{Eq:hppc}, where $T_r$ is replaced with $T_r+T_A$, with $T_A$ being the FDI attack on temperature readings of the room. Also, $\underline{T}_A$ and $\overline{T}_A$ are the lower and upper bounds of the attack, which are set as -5 and 5 degrees in this study. It is also noted that constraints \eqref{eql_room}-\eqref{eql_wall} are included in the attack model to ensure dynamics of the building are not violated after the attack. Fig.~\ref{fig.attack} depicts the results of attacks on the building with PI and MPC controllers.

\section{Case studies and simulation results}
The proposed framework is built on MATLAB/Simulink and the attack algorithm is designed using 'fmincon' nonlinear solver. The whole framework is carried out on an Intel Core CPU i9-13900KF processor
at 5.80 GHz and 64GB RAM. Here, two case-studies are performed to demonstrate the performance of the proposed control scheme and explore the impact of the attack on the whole system. To evaluate the tracking performance of the controller, the Mean Square Error (MSE) is used here and expressed as follow:
\begin{equation}
    MSE=\frac{1}{N}\sum_{i=1}^{N}(x_i-\hat{x}_i)^2
\end{equation}
where $x_i$ is the setpoint and $\hat{x}_i$ is the actual temperature.
\subsection{Daily normal working scenario}
\begin{figure}[t!]
\centering 
\includegraphics[width=0.9 \columnwidth]{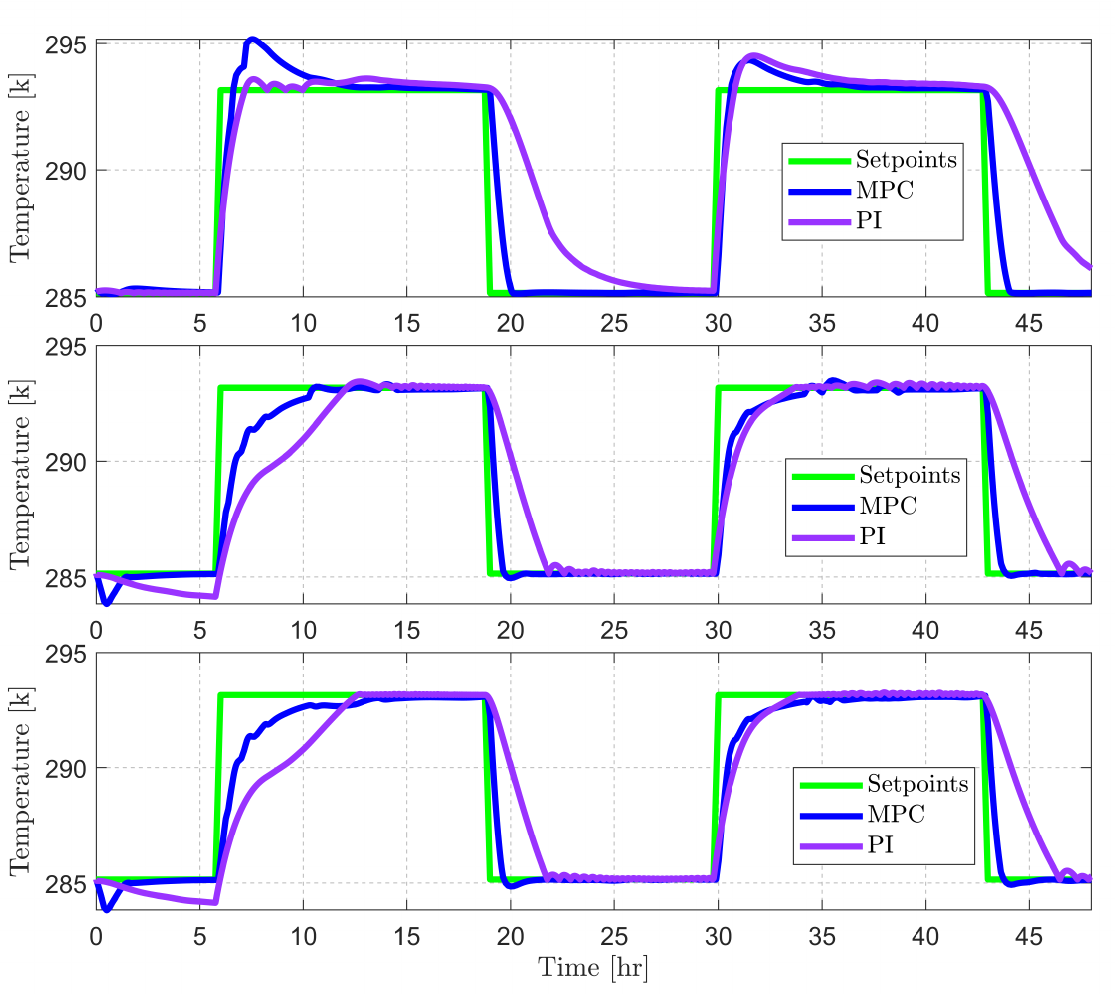}\vspace{-0.15in}
\caption{Comparative analysis of PI and MPC Controllers.}
\label{fig.centerwithoutattack}
\end{figure}

This case-study exhibits the performance of the proposed controllers under normal operation condition in winter. 
Fig.~\ref{fig.centerwithoutattack} illustrates a direct comparison of the PI and MPC controller outputs in relation to a predefined setpoint in different zones. Table.~\ref{tab:res} presents the numerical evaluation of different controllers. The results indicate superior performance of MPC compared to the PI controller under normal operational conditions. This aligns with our expectations, attributed to the prediction models in the MPC framework.

\begin{table*}[t!]
\centering
\caption{Numerical results for different controller under various operating condition}
\label{tab:res}
\begin{tabular}{cccccccccccc}
\hline\hline
\multicolumn{2}{c}{} & \multicolumn{3}{c}{MSE} & \multicolumn{3}{c}{Power Comsumption(kWh)} & \multirow{2}{*}{Lifespan (Year)} & \multicolumn{3}{c}{Valve State} \\ \cline{1-8} \cline{10-12} 
\multicolumn{2}{c}{Room}                                & Center & South  & West   & Center & South   & West    &       & Center & South & West \\ \hline
\multicolumn{1}{c|}{\multirow{2}{*}{PI}}  & Normal      & 5.3697 & 5.0636 & 5.0023 & 246.74 & 1542.16 & 1309.78 & 15    & 33     & 218   & 216  \\
\multicolumn{1}{c|}{}                     & With Attack & 5.8155 & 6.087  & 6.0901 & 237.74 & 1538.29 & 1312.73 & 15    & 980    & 144   & 220  \\
\multicolumn{1}{c|}{\multirow{2}{*}{MPC}} & Normal      & 1.5117 & 1.6813 & 1.6826 & 357.46 & 1626.35 & 1312.73 & 15    & 16     & 85    & 61   \\
\multicolumn{1}{c|}{}                     & With Attack & 2.3778 & 4.2815 & 4.5888 & 617.71 & 1818.00 & 1630.90 & 13.21 & 15     & 82    & 95   \\ \hline \hline
\end{tabular}
\end{table*}
\subsection{Room Temperature Sensor Attack}
\begin{figure}[t!]
\centering 
\includegraphics[width=0.95\columnwidth]{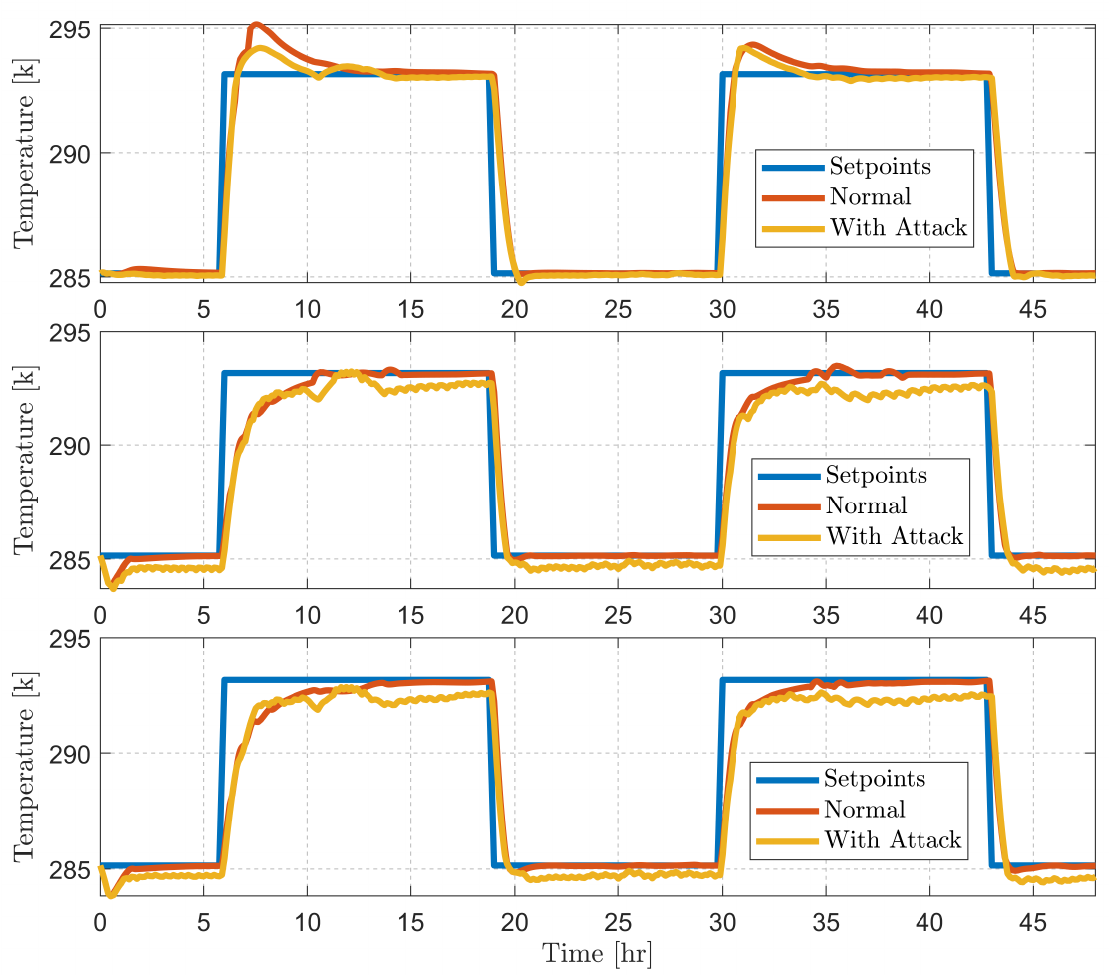}\vspace{-0.15in}
\caption{Center, south, and west rooms with MPC controller under attack.}
\label{fig:case2MPres}
\end{figure}

\begin{figure}[t!]
\centering 
\includegraphics[width=0.95 \columnwidth]{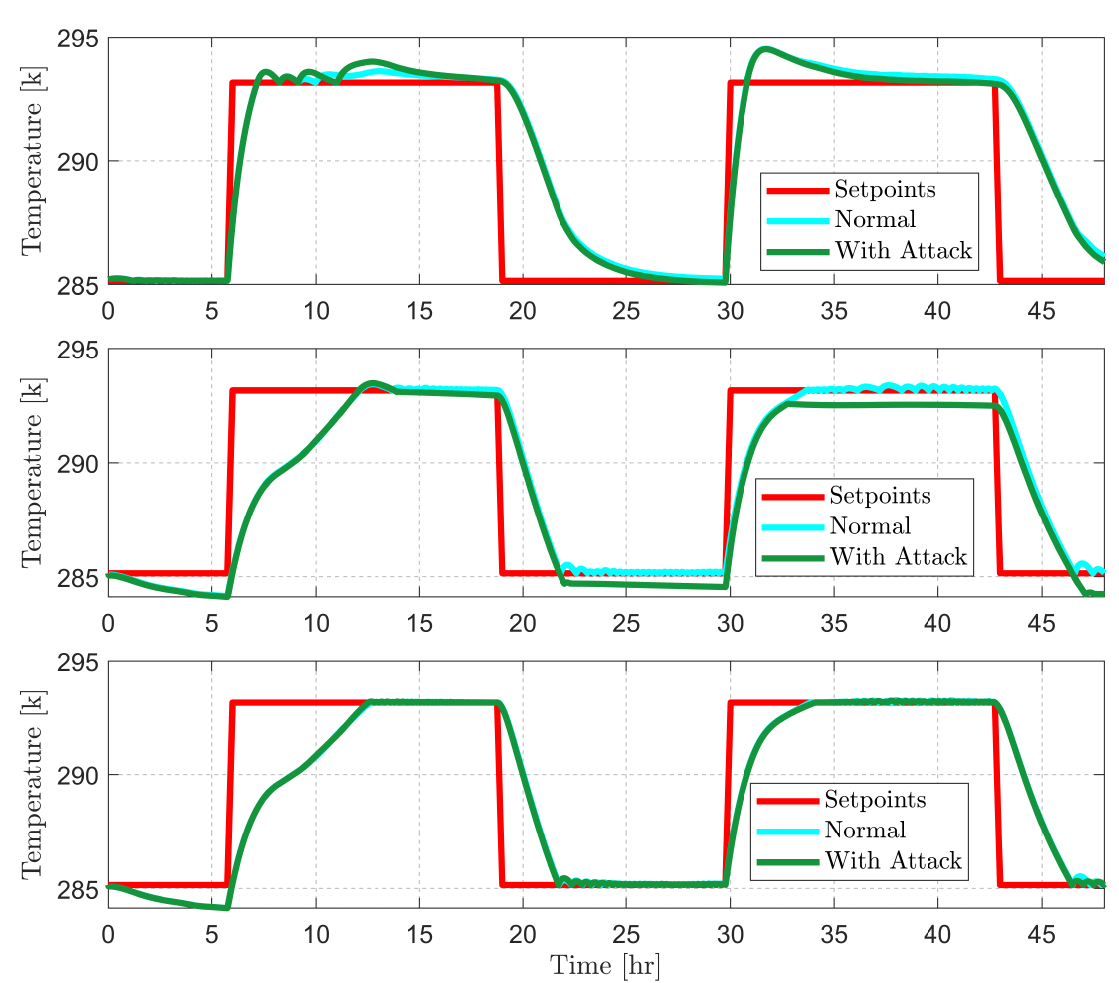}\vspace{-0.15in}
\caption{Center, south, and west rooms with PI controller under attack.}
\label{fig:case2pires}
\end{figure}

This case-study aims to explore the impact of room temperature sensor attack on the BMS. Based on our experiments, the south room temperature sensor attack was the most devastating one, therefore, we will focus exclusively on analyzing the cyber-attack targeting the south room temperature. The proposed attack algorithm is applied to both PI and MPC controllers. From Fig.~\ref{fig:case2MPres} - Fig.~\ref{fig:case2pires} and Table.~\ref{tab:res}, the proposed attack has a significant impact on the performance of both PI and MPC controllers. The resulting power consumption values are shown in Table.~\ref{tab:res} which illustrate the estimated heat pump power consumption (kWh) for 3 months. 
Analyzing power consumption under typical operational conditions and factoring in a 15-year lifespan, the results of the FDI attack demonstrate a more significant impact on the MPC controller. Specifically, the lifetime of the heat pump is reduced to 13.2 years after just 3 months of operation under the influence of the attack. However, as observed in the valve operation times in Table.~\ref{tab:res}, the PI controller results in reduced lifetime of valves as the valves operate 980 times during the attack for the center room compared with 33 times before the attack. The MPC provides enhanced valve protection by restricting changes to the valve state within the interval until the next sampling time, i.e., every 5 minutes. 


\section{CONCLUSIONS}

This paper has provided a detailed analysis of targeted cyberattacks on lifetime of assets in commercial buildings, focusing on the performance and resilience of PI and MPC controllers under normal conditions and cyber-attack scenarios. The findings show that MPC controllers outperform PI controllers in adaptability and efficiency, yet they are more vulnerable to reducing heat pump lifespan during cyberattacks. Conversely, PI controllers rapidly counteract such attacks, protecting the heat pumps but at the expense of reduced valve lifetimes.  The comparative analysis highlights the critical need for a robust control system in the face of evolving cybersecurity threats, which will be the focus of future research.

\bibliographystyle{IEEEtran}
\bibliography{IEEEabrv,conference_101719}

\begin{thebibliography}{10}
\providecommand{\url}[1]{#1}
\csname url@samestyle\endcsname
\providecommand{\newblock}{\relax}
\providecommand{\bibinfo}[2]{#2}
\providecommand{\BIBentrySTDinterwordspacing}{\spaceskip=0pt\relax}
\providecommand{\BIBentryALTinterwordstretchfactor}{4}
\providecommand{\BIBentryALTinterwordspacing}{\spaceskip=\fontdimen2\font plus
\BIBentryALTinterwordstretchfactor\fontdimen3\font minus \fontdimen4\font\relax}
\providecommand{\BIBforeignlanguage}[2]{{%
\expandafter\ifx\csname l@#1\endcsname\relax
\typeout{** WARNING: IEEEtran.bst: No hyphenation pattern has been}%
\typeout{** loaded for the language `#1'. Using the pattern for}%
\typeout{** the default language instead.}%
\else
\language=\csname l@#1\endcsname
\fi
#2}}
\providecommand{\BIBdecl}{\relax}
\BIBdecl

\bibitem{solano2021hvac}
J.~Solano, E.~Caama{\~n}o-Mart{\'\i}n, L.~Olivieri, and D.~Almeida-Gal{\'a}rraga, ``Hvac systems and thermal comfort in buildings climate control: An experimental case study,'' \emph{Energy Reports}, vol.~7, pp. 269--277, 2021.

\bibitem{lamnabhi2017systems}
F.~Lamnabhi-Lagarrigue, A.~Annaswamy, S.~Engell, A.~Isaksson, P.~Khargonekar, R.~M. Murray, H.~Nijmeijer, T.~Samad, D.~Tilbury, and P.~Van~den Hof, ``Systems \& control for the future of humanity, research agenda: Current and future roles, impact and grand challenges,'' \emph{Annual Reviews in Control}, vol.~43, pp. 1--64, 2017.

\bibitem{xie2022ga}
R.~Xie, T.~Zhang, X.~Jiao, and Q.~Yang, ``Ga optimized fuzzy pid control with modified smith predictor for hvac terminal fan system,'' in \emph{2022 IEEE 11th Data Driven Control and Learning Systems Conference (DDCLS)}.\hskip 1em plus 0.5em minus 0.4em\relax IEEE, 2022, pp. 1098--1104.

\bibitem{hussain2018internal}
S.~Hussain, S.~Gupta, and R.~Gupta, ``Internal model controller design for hvac system,'' in \emph{2018 2nd IEEE International Conference on Power Electronics, Intelligent Control and Energy Systems (ICPEICES)}.\hskip 1em plus 0.5em minus 0.4em\relax IEEE, 2018, pp. 471--476.

\bibitem{gupta2016controller}
S.~Gupta and R.~Gupta, ``Controller parameter optimization using hybrid bfoa-pso algorithm,'' in \emph{2016 IEEE 1st International Conference on Power Electronics, Intelligent Control and Energy Systems (ICPEICES)}.\hskip 1em plus 0.5em minus 0.4em\relax IEEE, 2016, pp. 1--5.

\bibitem{jiang2022building}
Y.~Jiang, S.~Zhu, Q.~Xu, B.~Yang, and X.~Guan, ``Building temperature and humidity adaptive control for a multi-zone hvac system using hybrid modeling method,'' in \emph{2022 IEEE International Conference on Industrial Technology (ICIT)}.\hskip 1em plus 0.5em minus 0.4em\relax IEEE, 2022, pp. 1--6.

\bibitem{fzy}
A.~Chojecki, M.~Rodak, A.~Ambroziak, and P.~Borkowski, ``Energy management system for residential buildings based on fuzzy logic: design and implementation in smart-meter,'' \emph{IET Smart Grid}, vol.~3, no.~2, pp. 254--266, 2020.

\bibitem{nn}
M.~Macarulla, M.~Casals, N.~Forcada, and M.~Gangolells, ``Implementation of predictive control in a commercial building energy management system using neural networks,'' \emph{Energy and Buildings}, vol. 151, pp. 511--519, 2017.

\bibitem{chen2022mbrl}
L.~Chen, F.~Meng, and Y.~Zhang, ``Mbrl-mc: An hvac control approach via combining model-based deep reinforcement learning and model predictive control,'' \emph{IEEE Internet of Things Journal}, vol.~9, no.~19, pp. 19\,160--19\,173, 2022.

\bibitem{ostadijafari2020tube}
M.~Ostadijafari and A.~Dubey, ``Tube-based model predictive controller for building’s heating ventilation and air conditioning (hvac) system,'' \emph{IEEE Systems Journal}, vol.~15, no.~4, pp. 4735--4744, 2020.

\bibitem{valenzuela2019closed}
P.~E. Valenzuela, A.~Ebadat, N.~Everitt, and A.~Parisio, ``Closed-loop identification for model predictive control of hvac systems: From input design to controller synthesis,'' \emph{IEEE Transactions on Control Systems Technology}, vol.~28, no.~5, pp. 1681--1695, 2019.

\bibitem{sheikh2019cyber}
A.~Sheikh, V.~Kamuni, A.~Patil, S.~Wagh, and N.~Singh, ``Cyber attack and fault identification of hvac system in building management systems,'' in \emph{2019 9th international conference on power and energy systems (ICPES)}.\hskip 1em plus 0.5em minus 0.4em\relax IEEE, 2019, pp. 1--6.

\bibitem{ramos2023lstm}
L.~S. Ramos and Z.~Yang, ``Lstm-based detection of ot cyber-attacks for an offshore hvac-cooling process,'' in \emph{2023 IEEE 6th International Conference on Electronic Information and Communication Technology (ICEICT)}.\hskip 1em plus 0.5em minus 0.4em\relax IEEE, 2023, pp. 943--948.

\bibitem{reda2022comprehensive}
H.~T. Reda, A.~Anwar, and A.~Mahmood, ``Comprehensive survey and taxonomies of false data injection attacks in smart grids: attack models, targets, and impacts,'' \emph{Renewable and Sustainable Energy Reviews}, vol. 163, p. 112423, 2022.

\bibitem{cyb1}
B.~K. Sethi, D.~Mukherjee, D.~Singh, R.~K. Misra, and S.~Mohanty, ``Smart home energy management system under false data injection attack,'' \emph{International Transactions on Electrical Energy Systems}, vol.~30, no.~7, p. e12411, 2020.

\bibitem{cyb2}
B.~K. Sethi, A.~Singh, S.~Mohanty, D.~Singh, and R.~Misra, ``Game theoretic smart residential buildings energy management system under false data injection attack,'' \emph{IEEE Internet of Things Journal}, vol.~10, no.~1, pp. 110--119, 2022.

\bibitem{zanetti2023performance}
E.~Zanetti, D.~Kim, D.~Blum, R.~Scoccia, and M.~Aprile, ``Performance comparison of quadratic, nonlinear, and mixed integer nonlinear mpc formulations and solvers on an air source heat pump hydronic floor heating system,'' \emph{Journal of Building Performance Simulation}, vol.~16, no.~2, pp. 144--162, 2023.

\end{thebibliography}

\end{document}